\documentclass{spie}

\usepackage{cite}
\usepackage{amsfonts} 
\usepackage{mathtools}

\usepackage{times} % Times New Roman
\usepackage{indentfirst} % Indent first paragraphs
\usepackage{algorithm}
\usepackage{algpseudocode}

\usepackage[colorinlistoftodos,color=pink]{todonotes} % For fancy \todo{something}
\usepackage{xkeyval}
\presetkeys{todonotes}{inline}{}

\usepackage{caption} 
\usepackage{lipsum} % For lorem ipsum, remove this

\usepackage[normalem]{ulem}
\useunder{\uline}{\ul}{}

\title{Portfolio Selection via Topological Data Analysis}

\author{Petr Sokerin\supit{1}, Kristian Kuznetsov\supit{1}, \skiplinehalf Elizaveta Makhneva\supit{1}, Alexey Zaytsev\supit{1}
  \skiplinehalf
  \normalsize 
  \supit{1}Skolkovo Institute of Science and Technology, Moscow, Russia
}

\begin{document}

\maketitle

\begin{abstract}

Portfolio management is an essential part of investment decision-making. However, traditional methods often fail to deliver reasonable performance. This problem stems from the inability of these methods to account for the unique characteristics of multivariate time series data from stock markets. We present a two-stage method for constructing an investment portfolio of common stocks. The method involves the generation of time series representations followed by their subsequent clustering. Our approach utilizes features based on Topological Data Analysis (TDA) for the generation of representations, allowing us to elucidate the topological structure within the data. Experimental results show that our proposed system outperforms other methods. This superior performance is consistent over different time frames, suggesting the viability of TDA as a powerful tool for portfolio selection.
  
\end{abstract}

\section{Introduction}

The stock market is an instrument through which shares of public companies can be bought and sold. It is a solid and influential economic instrument, and its understanding will lead to advances in financial study. Many studies consider the knowledge of the stock market, i.e. its underlying factors \cite{doi:10.1177/097265270500400302} or individual investor patterns \cite{Ngoc2013}. Stocks are divided into industries, which consist of companies that operate in different business areas. According to the Global Industry Classification Standard, 11 stock market sectors make up the entire stock market. It refers to the motivation for diversification of assets. Investors try to reduce risk and include internally independent stocks in their portfolio. However, the current categorization of stocks into economic sectors does not guarantee adequate diversification. For example, according to the standard categorization, Amazon is a retail company, but it's more related to the IT sector. One of the first studies on stock clustering showed a correspondence between sectors and the structure of stocks, although there are exceptions \cite{Mantegna1999HierarchicalSI}. We hypothesize that machine learning algorithms can re-cluster stocks more efficiently using the time series structure of the stocks. Our main goal is to compare different time series clustering approaches and evaluate them using financial metrics and portfolio valuation.

The ever-growing complexity of financial markets requires innovative, advanced methods for analyzing, interpreting, and predicting market trends. Traditional, conventional techniques sometimes fail to capture the intricate patterns in financial data. Recently, Topological Data Analysis (TDA) has emerged as a promising tool in data science, with the potential to unravel complex structures in high-dimensional datasets. TDA has the unique ability to understand the "shape" of the data, providing a novel perspective for understanding data patterns that often go undetected by traditional methods. The primary problem this research aims to address is to investigate the application of TDA for clustering stocks, thereby reducing investment risks. Specifically, this research: (1) introduces a novel approach for re-clustering sectors, (2) evaluates the potential of TDA-driven stock clustering to aid in portfolio optimization, risk management, and investment decision making, (3) compares the results to clustering using more traditional methods and modern machine learning techniques. We describe the current approaches and studies on financial clustering and the research gap; then we describe the pipeline and TDA essentials. Finally, we analyze our approaches over three time periods using U.S. stocks and compare them to baselines.

%Hence, our approach is to apply different algorithms to extract features from time series or get their embeddings. We use classical techniques (feature engineering), neural networks (such as classical LSTM), and more mathematical approaches like topological data science. As the final step, we provide the evaluation of the portfolio created according to each clustering approach and rank them by the financial metrics such as mean risk, average returns, and Sharpe ratio. 

\section{Related work}

There are several techniques for portfolio management using predictive models based on time series clustering. Generally, there are feature-based and distance-based approaches \cite{Aghabozorgi2015TimeseriesC}. In the feature-based approach, we extract a low-dimensional embedding vector from the time series and use it as a data point for the corresponding sample. In the distance-based approach, we use distance measures to construct the similarity matrix. In this study, we consider only feature-based approaches. One of the first works on financial time series clustering introduces the distance matrix obtained from the correlation matrix of stocks; they use a minimum spanning tree and then apply hierarchical clustering \cite{Mantegna1999HierarchicalSI}. Then many distance measures have been introduced based on linear predictive coding cepstrum \cite{Kalpakis2001DistanceMF}, PCA and Granger causality networks \cite{Billio2011EconometricMO}, or dynamic time wrapping \cite{10.1007/978-981-33-4501-0_3}.

\textbf{Feature engineering.} The most straightforward approach is to manually generate some features from the time series and use them as embeddings. The standard features are handcrafted statistics, value at risk, sharp coefficient, etc. \cite{Htun2023SurveyOF} Automatic feature extraction methods such as SAX \cite{Lin2007ExperiencingSA} or Bag-of-Features \cite{6497440} have been used over the years. One of the latest frameworks is Tsfresh, which combines many time series characterization methods and applies feature selection based on automatically configured hypothesis tests on top of \cite{Tsfresh}.

\textbf{Representation learning.} More advanced approach is to use deep learning to extract embeddings from stock time series \cite{CHONG2017187}. A variety of techniques from simple MLP \cite{ganatr2010spiking} to LSTM \cite{rajakumari2020forward} are used in studies. In general, we can use autoencoders to transform the original data into latent space with the encoder part of the model. Many implementations in media time series \cite{8066548} or stock clustering \cite{Tavakoli2020AnAD} confirm that autoencoders are useful for sequential data. Some general approaches to time series representation have been proposed. The main ones are contrastive learning framework TS2Vec \cite{ts2vec} and skip-gram method of representation learning Signal2Vec \cite{Signal2Vec}. Transformer models successfully used to learn speech patterns \cite{att_all_you_need} can also be applied to continuous sequential data in time series analysis \cite{wen2023transformers}. Transformers have been used for index prediction \cite{WANG2022118128} and stock movement prediction \cite{10.5555/3491440.3492080}.

\textbf{Topological Data Analysis (TDA).} Many researchers hypothesize that financial time series can be distinguished by some complex geometric features, which is a field of TDA \cite{Seversky2016OnTT}. It allows tracking changes in a structure across varying thresholds for different objects: scalar functions, point clouds, and weighted graphs. We can capture surface and structural patterns in high-dimensional data using TDA \cite{Chazal2017AnIT}. One of the main tools of TDA is persistence plots, or persistence barcodes, first described in \cite{Bar94}. Statistical features derived from barcodes can be used as a feature space to analyze graphs, point clouds, and so on \cite{ghrist2008barcodes}. In TDA for time series, the latter is usually obtained by time delay embeddings \cite{Packard1980GeometryFA, Takens1981DetectingSA}. 
The methods like persistence landscapes \cite{Bubenik2020ThePL} and persistence images \cite{Adams2017PersistenceIA} vectorize persistence diagrams in a more advanced way. Some papers use TDA to analyze financial time series. For example, quantified persistence landscapes have been used to detect early signals of market crashes \cite{tda_fin}. Time series clustering is also discussed in the literature with applications to spatial data \cite{topots2} and finance \cite{topots3}. The authors in~\cite{rivera2019topological, akcora2020bitcoinheist} applied TDA to portfolio management for cryptocurrencies. However, they used only basic features and studied a limited area of the cryptocurrency market, which is less competitive than a general stock market.

\textbf{Research gap.} There is still a need for a reliable approach to portfolio management for a stock market. The emergence of new techniques based on TDA suggests that we can generate meaningful feature representations for stock time series that capture their distinct internal structure. TDA has been applied to some areas of financial time series analysis, but has never been used to re-cluster stocks and perform portfolio selection. We believe that such a novel approach will be able to capture the complex dynamics of the stock market, making the model applicable to industry.

\section{Methods}

In financial markets, understanding the intricate, high-dimensional relationships between stocks is critical to strategic decision-making and portfolio optimization. However, methods such as conventional statistical clustering and dimensionality reduction techniques may prove inadequate to capture the complex, non-linear dependencies inherent in financial data. To address these shortcomings, our research focuses on constructing meaningful embeddings for stocks.
We also pay attention to the subsequent clustering using these embeddings. Embedding construction, or dimensionality reduction, transforms the high-dimensional stock data into a lower-dimensional representation while preserving the inherent structure and relationships of the data. Traditional methods such as Principal Component Analysis (PCA) can fall short because of the assumptions they make about the data. 
% For example, PCA assumes linearity in the data, which oversimplifies the model. 
Therefore, we propose to develop a novel approach based on machine learning to create more nuanced embeddings that can encapsulate the non-linear characteristics unique to financial data.

We define and compare procedures for extracting embeddings from time series. The general pipeline consists of embedding-based clustering and evaluation of selected embeddings via clustering. The pipeline for experiments includes the following steps:

\begin{enumerate}
    \vspace{-10pt}
    \setlength\itemsep{0em}
    \item \textbf{Getting embeddings from time series data}. We use different methods to get vector representation for every stock. 
    Our contribution is the development of topological methods. They are described in details below.
    We also consider several baseline methods based on feature engineering, classical machine learning, and neural network methods. 
    \item \textbf{Clustering}. We apply K-Means and Agglomerative clustering algorithms over the received on the previous step representations. The best clustering method for each case according to the risk metric is presented in the results. The clusters are used as an alternative to economic sectors for financial evaluation. We choose a number of clusters equal to $11$ as a number of economic sectors.
    % \item \textbf{Clustering evaluation}. The results of the clustering procedure are estimated by classical clustering metrics. We calculate the metrics using representation vectors and predicted cluster labels. Besides, we compare clustering results with real economic sectors using the Homogeneity score. 
    \item \textbf{Stock selection}. To build portfolio models, several stocks from each cluster were selected. We assume that our investor has a limited amount of money and can't build a portfolio of hundreds of stocks. We select the two best stocks from each cluster according to the Sharpe ratio calculated on the train set. After this step, we have a list of $22$ stocks for portfolio modeling. 

    % $$ Sharpe_i = \frac{return_i - return_{rl}}{\sigma_{i}} $$

    % where $return_i$ - mean return of stock $i$ for the train period, $return_{rl}$ - riskless return (we take a value annual 3\%), $\sigma_{i}$ - standard deviation of returns of stock $i$ for the train period

    \item \textbf{Portfolio backtesting}. The main goal is to get a vector of time series of portfolio price changing $\textbf{y}$ for the test period. To get more realistic results, we provide an evaluation with a backtesting procedure described below.
    \vspace{-10pt}
\end{enumerate}

\begin{figure}
    \includegraphics[width=1.0\linewidth]{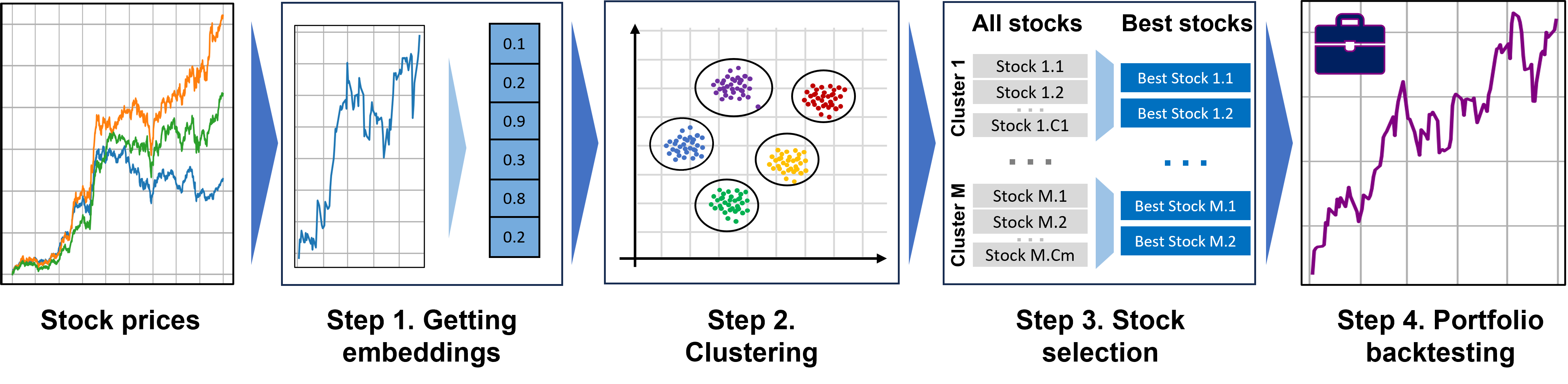}
    \caption{Clustering-based portfolio selection pipeline.}
    \label{fig:pipeline_portfolio}
\end{figure}

The pipeline is shown in 
Figure~\ref{fig:pipeline_portfolio}. After backtesting, we obtain a vector of daily returns for our portfolio. Based on this vector, we calculate the financial metrics, the mean annual risk, and the Sharpe ratio. In the following, we focus on step $1$, where we propose to create embeddings of time series using methods of topological data analysis. All other steps remain fixed between different methods, allowing us to evaluate the quality of the obtained embeddings within the whole pipeline.

\subsection{Embeddings from topological data analysis} 

\textbf{Preliminaries.} There are many fundamental tools in topology that can be used to describe universal properties of data, regardless of scale. These tools allow us to learn about the shape of the data, even in high dimensions. 
One of the main approaches is to track the changes of a structure across different thresholds for different objects, such as scalar functions, point clouds, and weighted graphs. 
TDA methods can be used to capture surface and structural patterns in high-dimensional data \cite{Chazal2017AnIT}. Below we describe the TDA essentials of working with point clouds and then how to obtain them from time series.

The classical approach is to construct a simplicial complex from the data. Following \cite{Edelsbrunner2009ComputationalT}, we denote $\sigma$ by a convex hull of a set of $k+1$ affinely independent points in $\mathbb{R}^d$ and call it $k$-simplex. The convex hull of the subset of these points is called the face of $\sigma$. Then a set of simplices is simplicial complex $K$ if: the face of $\sigma \in K$ is also in $K$ and the non-empty intersection of $\sigma_1, \sigma_2 \in K$ is the face of both. The special case of the simplicial complex is the Vietoris-Rips complex, which is commonly used in topological data analysis. It can be defined from any metric space and is useful for many practical tasks. The Vietoris-Rips complex $\text{VR}_{r}(P)$ is defined on the metric space $(P, d)$ for $r$ and consists of all subsets of diameter at most $2r$: $\sigma \in\text{VR}_{r}(P) \iff \forall p,q \in \sigma, d(p, q) \leq 2r$. Since the complex can be computed from the point cloud, we can also analyze its homology. The $p$-th homology group $\mathsf{H}_p$ represents the $p$-dimensional "holes". For example, $\mathsf{H}_0$ represents connected components and $\mathsf{H}_1$ represents basic cycles. However, we can only obtain this topological information for the simplicial complex that we derive from the data for a given threshold. Persistence diagrams, or barcodes, first described in \cite{Bar94}, allow us to obtain information for the entire range of thresholds, eliminating the need for this hyperparameter. Considering the range of values $r_0 < r_1 < ... < r_n$ for the Vietoris-Rips complex construction, we obtain an increasing sequence of subcomplexes called $\mathbb{R}$-filtration: $ \emptyset \subseteq K_{r_0} \subseteq K_{r_1} \subseteq ... \subseteq K_{r_n} $. The filtration allows to obtain the basis for the subspaces of the filtration. Converted into "canonical form" it gives information about the thresholds for birth and death of any member of a homology group. This can be seen as a collection of intervals.

\begin{figure}
    \centering
    \includegraphics[width=0.7\linewidth]{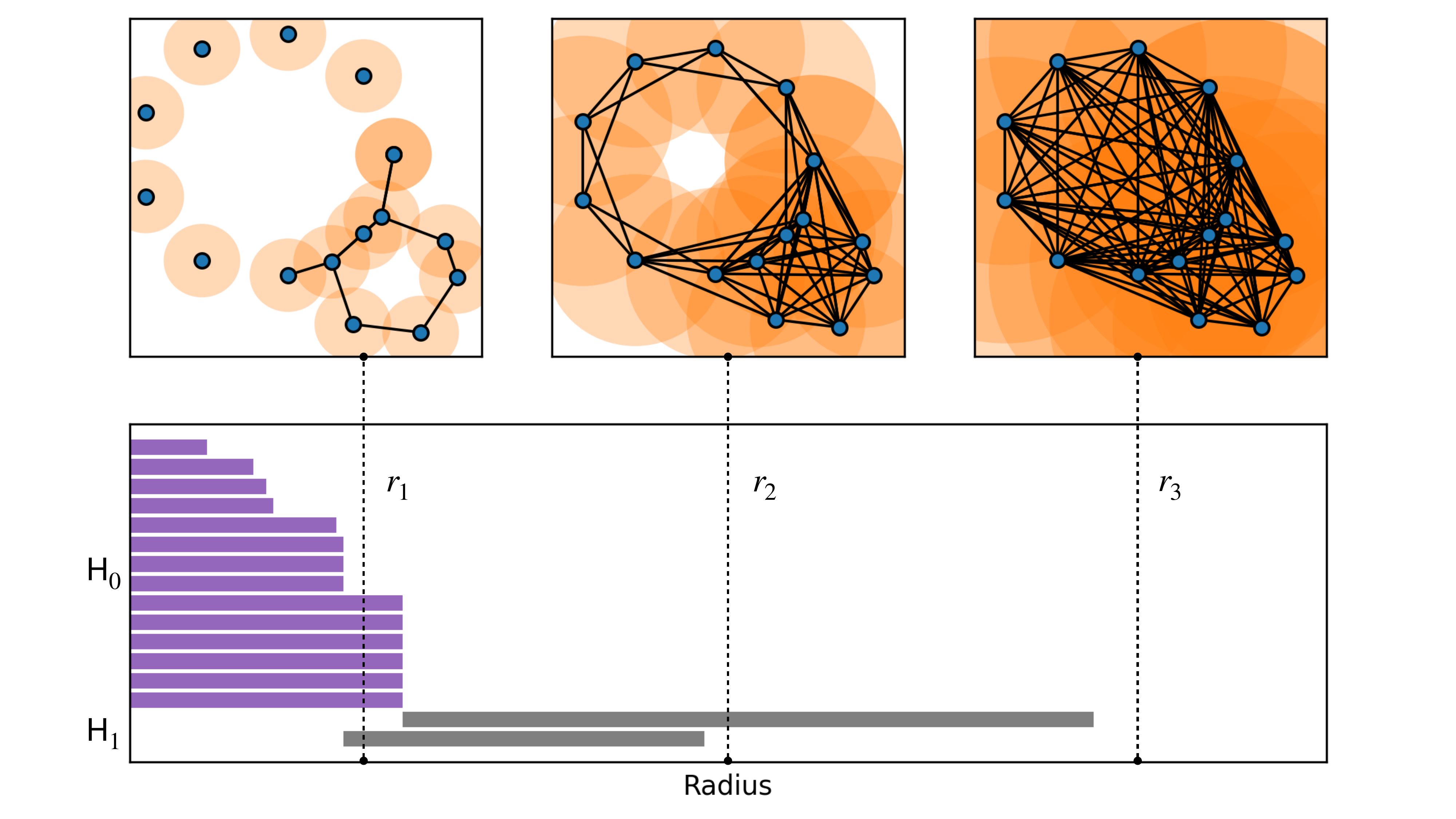}
    \caption{Example of the persistence barcodes.}
    \label{fig:barcodes}
\end{figure}

\textbf{TDA for time-series.} The general approach to time series analysis from a TDA perspective is to map a time series to a point cloud. The mapping is done by time delay embeddings \cite{Packard1980GeometryFA, Takens1981DetectingSA}. They translate each point in a $1$-dimensional time series into a $d$-dimensional space by picking the point from a sliding window of fixed size. For each $t_i$ in the time series $T = \{t_i\}_{i = 1}^N$, we extract a point $[t_i, t_{i + \tau}, t_{i + 2\tau}, \ldots,t_{i + (d-1)\tau}] \in \mathbb{R}^d$, where $d$ is the embedding dimension, $\tau$ is the time delay, and $(d - 1)\tau$ is the size of the sliding window. The stride can be introduced to reduce the number of points sampled from the time series. We use this procedure to obtain the point cloud $M(T)$ from each stock time series $T$. To choose the optimal $d$ and $\tau$, we first minimize the time delayed mutual information between different values of $\tau$. Then the embedding dimension is chosen by the approach proposed in \cite{kennel1992determining}. We perform this procedure for each stock to construct a set of point clouds for the collection of stocks. Then for each stock we obtain the persistence barcodes.

The study considers 5 approaches to extract the embedding from the persistence barcodes. The first one is just a collection of statistical features based on the barcodes. For all dimensions of the barcodes, we compute the mean, sum, standard deviation, and maximum of the lengths. For all dimensions greater than $0$, we also compute the last birth threshold, the birth and death of the longest bar, and the number of bars. We will call this Bars statistics. Other approaches are vectorizations of persistence diagrams: persistence landscapes \cite{Bubenik2020ThePL} and persistence images \cite{Adams2017PersistenceIA}. In the first method, we rotate the diagram 45 degrees to make it easier to proceed, and then introduce a set of functions. These functions are isosceles triangles with points from our diagram at the top. We take the maximum of this set of functions and get the first-layer persistence landscape function. The second is the method that uses the function of 2 variables: birth and death. This function at each point is the product of the piecewise linear weight function and the normalized symmetric Gaussian. By discretizing this function, a persistence diagram is transformed into a persistence image that can be flattened into a vector representation. We use $1$- and $2$-dimensional persistence landscapes and persistence images as embeddings, denoted PL1, PL2 and PI1, PI2, respectively.

\section{Results}

\subsection{Evaluation}
As data for time series clustering, we use the market prices of the best American stocks from the Standard and Poor 500 index (S\&P 500). We select three time series periods for estimation. The first period \textbf{USA 2012} contains stock price data for the period 2012-2013 as a train and 2014-2017 as a test; the second period \textbf{USA 2015} -- 2015-2016 as a train and 2017-2019 as a test; the third period \textbf{USA 2018} takes 2018-2019 as a train and 2020-2022 as a test. For all datasets, the embeddings and clustering were only provided for training step. The test data was used for financial portfolio backtesting and result estimation. 
As basic economic sectors of the stocks, we take 11 sectors from the official classification in the U.S.: Healthcare, Industrials, Consumer Cyclical, Technology, Consumer Defensive, Utilities, Financial, Basic Materials, Real Estate, Energy, and Communication Services. 

% We will use two main groups of metrics for results evaluation: clustering and financial. The results of the clustering procedure are estimated by classical clustering metrics. We calculate the Silhouette score and Calinski-Harabasz index using embeddings and predicted labels.

To obtain financial metrics, we use the Markowitz portfolio model, minimizing investment risk and backtesting with a rolling window of one month for the test period. The backtesting procedure imitates portfolio rebalancing. We want to change our portfolio weights in each window test period because the market situation changes and our stock weights become obsolete. In the end, we get a time series of the portfolio price change for the test period $y$. We choose the mean annual risk of the portfolio and the Sharpe ratio as the main financial metrics. We want to minimize the mean annual risk, which can be calculated as $\sigma(y) \sqrt{252}$, where the normalization coefficient $252$ is the approximate number of working days in a year. The Sharpe ratio is the ratio of return to risk. The higher the ratio, the better. 

Dividing stocks into appropriate categories can help with portfolio selection. An example of such an approach is using basic economic sectors to label stocks. These \textbf{Economic sectors} allow one to avoid using machine learning and expert-based clustering to select the portfolio. We compare this strong baseline to several clustering approaches based on extracting the embeddings from the percentage changes of stock prices over a given time period. Other baselines can be established for feature engineering from stock price changes. They are called \textbf{classical features}. We apply two baselines: using financial features and using an automatic feature extraction method TSFresh \cite{Tsfresh}. We compare these two approaches and store only financial features because of their better performance. We also apply \textbf{Dimensionality reduction} methods: PCA, UMAP and FastICA. The last one shows better results, so we use this approach. 

The other three groups of basic methods are based on neural networks. These groups are \textbf{Autoencoders}, \textbf{Transformers}, and \textbf{Neural network embeddings}. As autoencoders, we have tried several architectures: LSTM-based, CNN-based, and Fully-Connected models, but the LSTM-based model showed better results than other autoencoders. The main idea of the transformer approach is to use the outputs of different transformer layers as embeddings. We take the mean output of the embedding layer and all encoder layers as representation vectors for clustering. 
As neural network embeddings, we used the Signal2Vec \cite{Signal2Vec} model, which is similar to the classical Word2Vec but can be used on continuous data. We try to use the TS2Vec model \cite{ts2vec}, which is based on self-supervised contrast training with 1D convolutional layers, but it performs worse than Signal2Vec.  Financial metrics are also compared to the financial index S\&P 500. 
This index represents market behavior and is traditionally used as the main benchmark in finance. 
The index contains about 500 stocks, and it's a complicated task to achieve comparable results in risk and return with any stock selection method. However, we outperform it in some other experiments. 

% \subsection{Clustering quality}
% To estimate the quality of the clustering, we use two metrics: Silhouette Score and Calinski Harabasz (CH) -- we want both to be higher. For all periods we have Signal2Vec as the best model for CH, while for Silhouette Score the results are different for different periods. However, from a CH point of view, Signal2Vec also provides comparable quality. As for other models, in the period 2012-2017, the best model is a transformer with embedding as the sum of the output of the embedding layer and the hidden states of the layers in the encoder. In the years 2015-2019, the autoencoder with LSTM layers shows the best clustering according to CH as well as in the period 2018-2022.

% \input{tables/clustering}

\subsection{Portfolio quality}

In Table ~\ref{table:category} we show financial metrics for different periods. The best algorithm according to our main metric, risk, is the TDA method using the PI1 method. It is in second place for the period 2012-2017, after the S\&P 500 Index (Index), which is in first place. Interestingly, there is a large gap between the metric of our model and all the others. It seems that the TDA model captures hidden properties of the stock time series related to their behavior. However, it is very close to the S\&P 500 in all periods, even though the index uses 500 stocks for the strategy while our portfolio uses only 20. Our method is therefore able to achieve lower risks with lower costs. In 2015-2019 and 2018-2022, our model has the lowest risk of all the baselines. For the 2015-2019 period, only our model and the index were able to reduce the portfolio's risk through diversification. It is worth noting that the Sharpe ratio of our model is several times higher than that of the index.
Thus, our model is able to provide higher returns despite the lower risk. Nevertheless, TDA method is not one of the best algorithms according to the Sharpe ratio, except for the years 2012-2017. However, this is a normal situation because we don't want to achieve the best risk-return ratio, we want to reduce the risk and have a comparable level of Sharpe with the index. So we achieve our main objectives. The methods with a better Sharpe ratio have a higher risk than TDA. 

% Approaches with vectorized time series, Signal2Vec, and TS2Vec, didn't show good clustering results. We also see that both models rather reduce risk or increase returns. However, aggregated metrics Sharpe indicate that increase in returns for TS2Vec is bigger than the increase of the risk, so on average portfolio constructed via this technique is better than the baseline but this wasn't the purpose.

We also want to take a closer look at other methods. 
The financial features approach provides one of the best results according to the Sharpe ratio, and its risk is lower than the economic sectors approach in the period 2012-2017. This is one of the best models for this period, as this strategy allows both minimizing risk and increasing returns. However, it is not the best at all for 2018-2022, and it fails in the 2015-2019 period.
The transformers show a quality that is not as good as the best models according to risk metrics. This model also didn't manage to reduce risk compared to classical sector diversification in 2012-2017 and 2015-2019, but it worked in 2018-2022. The reason for this could be that the time series are similar in the row values, which could be confusing since we don't use tokens in our transformer. Also, the transformer wasn't built to differentiate the time series. It was trained to predict the next value in the series. Nevertheless, the transformer has a higher Sharpe ratio, and in 2015-2019 it was the best model according to this metric.

The LSTM autoencoders, FastICA dimensionality reduction model, and Signal2Vec also appeared unable to reduce risk or increase Sharpe ratio compared to the main baseline in 2012-2017 and 2015-2019.

%However, all models worked quite well in 2018-2022 according to risk metric. The best model from Sharpe's ratio point of view is a self-supervised model based on convolutional layers.

\begin{figure}[!h]
    \begin{center}
        \includegraphics[width=0.9\linewidth]{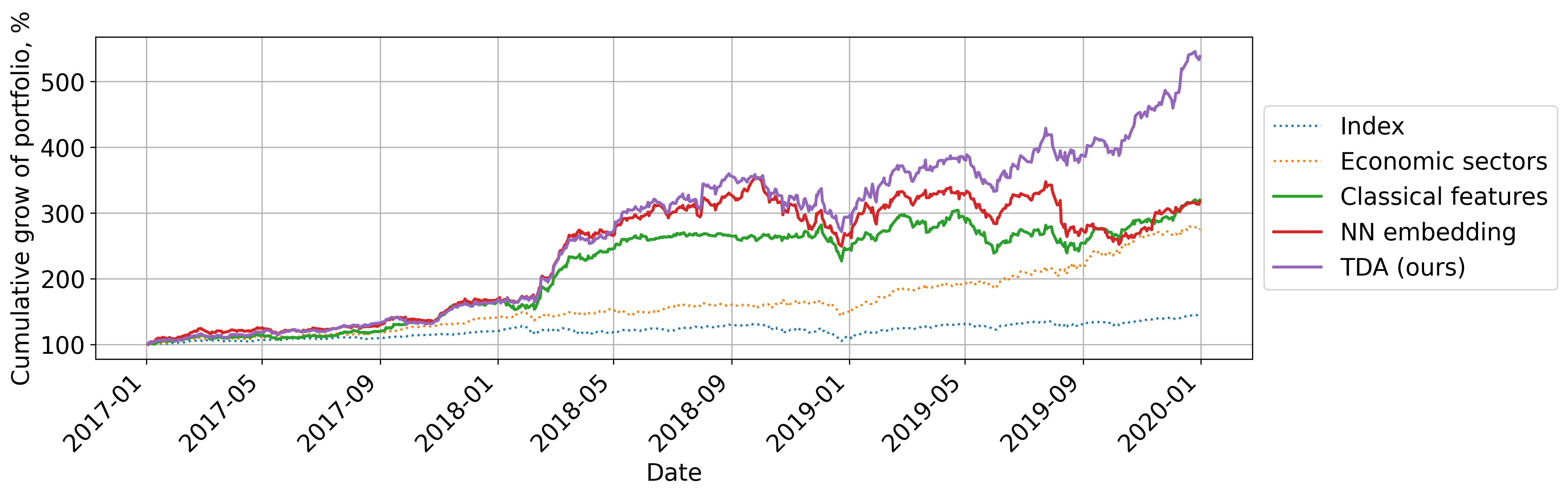}
    \end{center}
    \caption{Portfolio evaluation for the best models}
    \label{fig:portfolio}
\end{figure}

We also calculated the cumulative growth of the portfolios for the 2015-2019 dataset and the 2017-2019 test period. As shown in Figure~\ref{fig:portfolio}, our method allows to achieve the highest return from the available stocks at the end of the period, while having the lowest risks. This means that the topological method works quite well for portfolio diversification. All in all, we must emphasize that despite different risk approaches, some algorithms show not only lower risk but also better Sharpe ratio in comparison with the economic sector approach.

\begin{table}[]
\centering
\begin{tabular}{llcccccc}
\hline
\multicolumn{1}{c}{\textbf{Category}} & \multicolumn{1}{c}{\textbf{Method}} & \multicolumn{2}{c}{\textbf{USA 2012}} & \multicolumn{2}{c}{\textbf{USA 2015}} & \multicolumn{2}{c}{\textbf{USA 2018}} \\
\multicolumn{1}{c}{\textbf{}}         & \multicolumn{1}{c}{\textbf{}}       & \textbf{Risk}     & \textbf{Sharpe}   & \textbf{Risk}     & \textbf{Sharpe}   & \textbf{Risk}     & \textbf{Sharpe}   \\ \hline

Index S\&P500                                 & -                            & \textbf{0.1420}   & 0.0197            & {\ul 0.1363}      & 0.0436            & {\ul 0.2892}      & 0.0149            \\
Economic sectors                      & -                           & 0.2049            & 0.0795            & 0.1661            & {\ul 0.1355}      & 0.4155            & 0.0426            \\
Classical features                    & Financial features                      & 0.1853            & {\ul 0.084}       & 0.1768            & 0.1325            & 0.392             & 0.0514            \\
Dimensionality reduction              & FastICA                             & 0.2798            & 0.0444            & 0.2256            & 0.078             & 0.3613            & 0.0374            \\

Autoencoders                          & LSTM                                & 0.2376            & 0.0655            & 0.2190            & 0.1214            & 0.3992            & \textbf{0.0680}   \\

Transformers                          & All layers embeddings                              & 0.2116            & 0.077             & 0.1601            & \textbf{0.1589}   & 0.3934            & {\ul 0.0661}      \\

NN embeddings                       & Signal2Vec                          & 0.2016            & 0.0707            & 0.1790            & 0.081             & 0.4227            & 0.0399     \\ 
TDA features (ours)                   & PI1                                 & {\ul 0.1431}      & \textbf{0.1168}   & \textbf{0.1338}   & 0.0716            & \textbf{0.2780}   & 0.0304            \\
\hline     
\end{tabular}
\caption{Financial metrics Risk and Shape for different representation methods}
\label{table:category}
\end{table}

\subsection{Comparison of TDA approaches} 

\begin{table}[]
\centering
\begin{tabular}{lcccccc}
\hline
\multicolumn{1}{c}{\textbf{Method}}                      & \multicolumn{2}{c}{\textbf{USA 2012}}                                  & \multicolumn{2}{c}{\textbf{USA 2015}}                                  & \multicolumn{2}{c}{\textbf{USA 2018}}                                  \\
\textbf{}                      & \multicolumn{1}{c}{\textbf{Risk}} & \multicolumn{1}{c}{\textbf{Sharpe}} & \multicolumn{1}{c}{\textbf{Risk}} & \multicolumn{1}{c}{\textbf{Sharpe}} & \multicolumn{1}{c}{\textbf{Risk}} & \multicolumn{1}{c}{\textbf{Sharpe}} \\ \hline
Index S\&P500             & \textbf{0.142}                    & 0.0197                              & {\ul 0.1363}                      & 0.0436                              & {\ul 0.2892}                      & 0.0149                              \\

Economic sectors    & 0.2049                            & 0.0795                              & 0.1661                            & 0.1355                              & 0.4155                            & 0.0426                              \\ 

PI1    & {\ul 0.1431}                      & \textbf{0.1168}                     & \textbf{0.1338}                   & 0.0716                              & \textbf{0.278}                    & 0.0304                              \\ 
 
PI2     & 0.1539                            & {\ul 0.0932}                        & 0.1475                            & 0.1362                              & 0.3047                            & 0.0385                              \\ 
Bars statistics   & 0.2458                            & 0.0566                              & 0.2586                            & \textbf{0.1565}                     & 0.3851                            & {\ul 0.0641}                        \\ 
PL1 & 0.295                             & 0.0577                              & 0.2975                            & 0.1405                              & 0.4055                            & 0.0599                              \\ 

PL2 & 0.2414                            & 0.0797                              & 0.2539                            & {\ul 0.1545}                        & 0.4422                            & \textbf{0.0911}  \\
\hline
\end{tabular}
\caption{Financial metrics Risk and Shape for various TDA methods and Baseline Index S\&P 500}
\label{table:topology}
\end{table}

We show the financial metrics for all TDA approaches in the table~\ref{table:topology}. The persistence image (PI) is generally better than the persistence landscape (PL), and the bar statistics show unstable results. The advantage of PI is more stable representations with less weight given to noisy patterns, which is a common difficulty when working with financial time series \cite{Adams2017PersistenceIA}. In addition, PI is in Euclidean space, while PL must be sampled to obtain the embedding. This is another reason for the instability of PL embeddings. 

The minimization of risk is seen for the clustering based on one-dimensional barcodes related to the cyclical structures in the time series. PL1 captures the structure of both persistent and non-persistent cycles in the vector representation. This correlates with the notion of economic cycles that occur in financial time series. 
Notable examples are political \cite{pastor2020political} and business \cite{chauvet1999stock}. We hypothesize that one-dimensional barcodes are able to distinguish different categories of stocks by analyzing the deeper structure of economic cycles that cannot be captured by classical methods. The PI1 shows that diversification based on this information is indeed useful to minimize risk.
We believe that further analysis of one-dimensional barcodes can provide more stable and interpretable results.

% Among classical time series clustering methods we don't see really good results besides TSKMeans. We suppose that it is due to the distance-based approach of this method -- it may give us good clusters. But it is not necessarily beneficial for reducing risks. And we see it from financial metrics: here TSKMeans provide worse results in risks and returns than other methods and than the baseline. k-Shape has shown comparable to baseline results according to Sharpe.

% The same situation we see with dimensionality reduction techniques. We got noisy data with similar values which may result in mixed clusters. Interestingly, UMAP showed the lowest risk compared to all other models. All dimensionality reduction techniques show lower risks but with lower returns.

% Simple models based on manual features have shown not bad clustering results, better than some of the more complicated models. Via these methods, we may extract relevant features, and this way we don't forget about data structure. But still, we may miss important information and lose in quality. That is what we see from financial metrics: both risks are higher and returns are lower.

\section{Conclusion}

We have developed a method for applying Topological Data Analysis (TDA) to stock clustering, presenting a promising way to navigate the complexities of financial markets. 
The study proposes the most efficient use of TDA-based embeddings and provides an explanation as to why the particular method chosen is superior to others.
% Our research proves its practical advantages in the portfolio management.
% We proposed novel topological features for this task, rooted in the unique ability of TDA to explore and capture the multidimensional, non-linear relationships embedded within financial data. The efficacy of these features has been rigorously evaluated through our experiments, with results underscoring their superiority over traditional methods.
Our approach outperforms alternative strategies. We achieve the lowest or second lowest level of risk and a better risk-return ratio than the financial index. This can be attributed to our method's more comprehensive and accurate stock clustering, which provides a deeper understanding of market dynamics and thus informs more strategic asset allocation and risk management decisions. However, it is also important to note that while TDA-based stock clustering has proven to be powerful, it is only one piece of the broader financial analysis landscape. 

% In conclusion, using TDA to cluster stocks opens up a new paradigm in financial data analysis, offering the potential for more insightful and profitable investment decision-making. Our research contributes to this field, paving the way for a more nuanced understanding of financial markets and more effective portfolio management strategies.

\acknowledgements

This work was supported by the Russian Science Foundation (project 21-11-00373) grant.

\bibliographystyle{spiebib}
\bibliography{bibliography}

% This table is required by ICMV for review and will not be published

% \clearpage

% \section*{AUTHORS' BACKGROUND}

% \begin{table}[h]
% 	\centering
% 	\begin{tabular}{ | c | c | p{45mm} | c | } \hline
% 		Name & Title & \centering Research Field & Personal website \\
% 		\hline
% 		  Petr Sokerin & Master Student. & Data Science. Machine learning. Finance &   \\
% 		\hline
% 		Kristian Kuznetsov & Master Student & Natural Language Processing. Topological data analysis. & \\
%  		\hline
%  		 Elizaveta Makhneva & Master Student & Machine learning. Recommender systems & \\
%  		\hline
%  		Alexey Zaytsev & Assistant Professor & Deep learning. Machine learning. Statistics  & https://faculty.skoltech.ru/people/alexeizaitsev \\
% 		\hline
% 	\end{tabular}
% \end{table}

\end{document}